# Ordered carbon nanotubes and globular opals as a model of multiscaling photonic crystals

G. Dovbeshko[1], O. Fesenko[1], V. Moiseyenko[2], V. Gorelik[3], V. Boyko[1] and V. Sobolev[4]

[1]*Institute of Physics, NAS of Ukraine, 46, prospect Nauky, 03028 Kyiv, Ukraine*
[2]*Dnipropetrovsk National University, 72, prospect Gagarina, 49050 Dnipropetrovsk, Ukraine*
[3]*P.N. Lebedev Physics Institute Russian Academy of Sciences, 53, Leninsky prospect, Moscow 117924, Russia*
[4]*Technical Centre of NASU, 13, Pokrovskaya str., 04070 Kyiv, Ukraine*

**Abstract.** Experimental data on carbon nanotube carpet and synthetic opals have been compared by visible, infrared spectroscopy, and electron microscopy. Spectral features of the objects under study in infrared region are registered. Three regions of abnormal behavior of reflectance and absorbance for carbon nanotube carpet and two regions for opals in the 7000-2000 cm$^{-1}$ are observed in comparison with the separated nanotubes or SiO$_2$ globules in disordered states and/or condensed state. The spectroscopic features of the photonic crystals caused by their different microstructures at different length scales and basis for development of a proper model for light propagation through the photonic crystals in the IR region are presented for discussion.

## 1. Introduction

Photonic crystals as a new class of materials take place between nano- and microstructural highly ordered composites. Electromagnetic wave propagation through a structure with sizes of elements close to the wavelength leads to Bragg diffraction, multiple scattering, *etc*. As a result of this process, a forbidden photonic band arises [1-5]. Photonic crystals with a tunable forbidden band are of interest for researchers and technical applications. Our paper is directed to optical study of photonic crystals – opals and carbon nanotubes in IR region. Optical characteristics of opal were not studied in this region due to the fact that opal is a crystal that is traditionally applied in the visible region [5], however the SiO$_4$-structural element of SiO$_2$ single crystal has transparency windows in the IR region [6]. Carbon nanotube is a structure that absorbs light in a wide wavelength range from UV, through VIS and up to FIR region. The property of ordered nanotubes ensembles as a possible model of photonic crystals are not studied, too. Particularly, assignment of a nanotube carpet to photonic crystals should be proved. Our goal was to register the optical characteristics of opals and carbon nanotubes in IR region that are related with different size scales of microstructures and connected with the nature of crystalline material (phonon modes) and do a step for its modeling.

## 2. Methods and materials

The samples of carbon nanotubes (CNT) were obtained by evaporating either a solid carbon precursor (camphor) or a liquid one (cyclohexanol) on silicon substrates. The process of synthesis involves the coevaporation of the carbon precursor and ferrocene, used as the catalyst source, in nitrogen atmosphere [7]. The nanotubes were well packed, vertically aligned, multi walled and several millimeters in length.

Synthetic opals were produced in Zelenograd (Russia) and Dnepropetrovsk (Ukraine). Nanodisperse silica globules were synthesized by the method of Stober *et al.* through the hydrolysis of tetraethoxysilane Si(OC$_2$H$_5$)$_4$ in water-ethanol solution in the presence of ammonium hydroxide as catalyst. The molar ratio of components in the reaction mixture was as follows: NH$_4$OH:H$_2$O:C$_2$H$_5$OH:Si(OC$_2$H$_5$)$_4$=(0.1-1):(2-20):(11-14):0.14.

Structure of the samples and their optical properties has been characterized by SEM analysis and FTIR reflectance and absorbance spectroscopy, respectively. IR spectra have been registered with Bruker IFS-66 FTIR spectrometer in the 7500-400 cm$^{-1}$ region. SEM images of opals were obtained with EPMA SEI JXA-8200 microscope and carbon nanotubes images – with JSM-35 microscope.

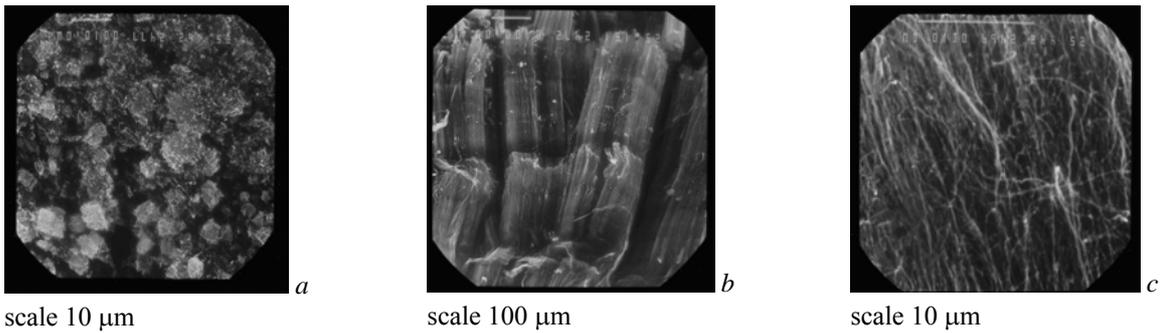

scale 10 μm  scale 100 μm  scale 10 μm

**Fig. 1.** Microstructure (SEM images) of carpet of carbon nanotubes consisting of separated column or blocks: *a* – top view, *b* and *c* – side view.

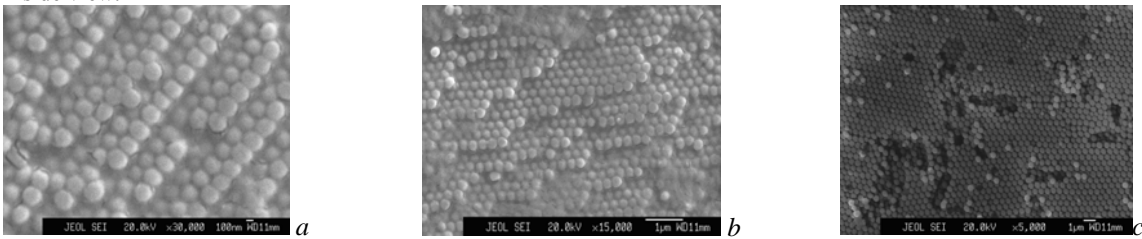

**Fig. 2.** Microstructure (SEM images in SEI mode) of synthetic opal: *a* – ordered globules (view from the top); *b* – globules organized in terraces of mosaic structure – "strip" type; *c* – disordering in a microstructure of opal (view from the top).

### 3. Results and discussion

According to M. Barabanenkov [1] and V. Kosobukin [2], a definition of photonic crystals or existence of forbidden photonic band (FPB) could be based on the following features: periodicity in three dimensions; small absorbance (multiple scattering); dielectric contrast $n_1/n_2 > 2/3$; a volume ratio of structural units of a photonic crystal to pore filling environment should be small; structure is good when optical path in both materials is the same; deviation in sizes is much less than the average particle size: $\Delta R \ll R_{avg}$; displacement of the structural elements is much less than a distance between the elements of the structure: $\Delta u \ll a$.

Taking into account the abovementioned criteria and dielectric constants for silica and graphite materials [8], we could suppose from SEM image analysis that carpet of nanotubes (Fig. 1) and synthetic opal (Fig. 2) could be assigned to photonic crystals. For photonic crystals, a simple equation for Bragg diffraction could be written as: $\lambda_{max} = 2d_{hkl} n_{eff} \sin\theta$, where $n_{eff} = [(n_s)^2 f + (n_p)^2 (1-f)]^{1/2}$, $d = (2/3)^{1/2} a$, $a$ – characteristic size (diameter) of the structural element of photonic crystal, $d$ – *d*istance between elements, $f = 0.74$ [3]. CNT carpet has a microstructure of a different scale – from micrometers to millimeters. For example, a characteristic size of structure granularity is 0.05-0.1 μm (Fig. 1a, c), 2-10 μm (Fig. 1a, c), and 1000-2000 μm (Fig. 1b). Similar situation is observed with opal having the fine globular structure with 0.2-0.5 μm size (Fig. 2a), above the globular structure (2-5 μm) (Fig. 2b, c) and superfine structure consisting of small balls with 0.05 μm size [9]. So, we suppose that photon stop bands could exist in visible and IR regions up to far-IR in the objects under study, if the main conditions for FPZ will be valid. The spectra presented in Figs 3 to 7 are an evidence of this asumption. Near-normal incidence IR reflectance spectra of the CNT carpet (Fig. 4a, b) and opal photonic crystal (Fig. 5) depend on the sample orientations.

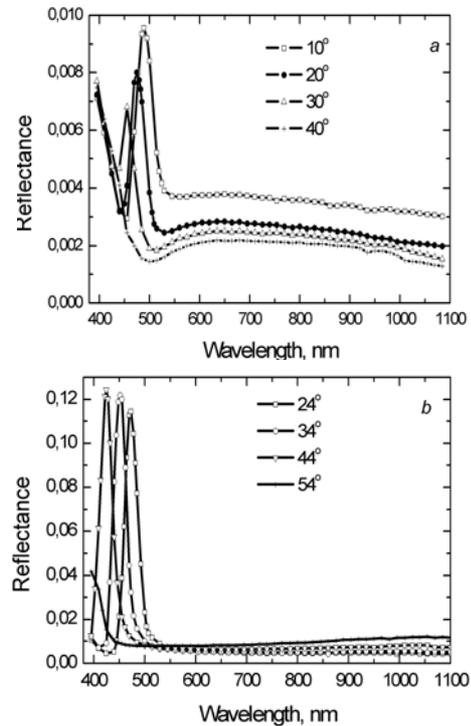

**Fig. 3.** Bragg reflectance of synthetic opal for two orthogonal orientations of the crystal in (111) plane at different angles of light incidence: *a* – big "axis" of the sample is parallel to the long axis, *b* – small "axis" of the sample is perpendicular to the short one.

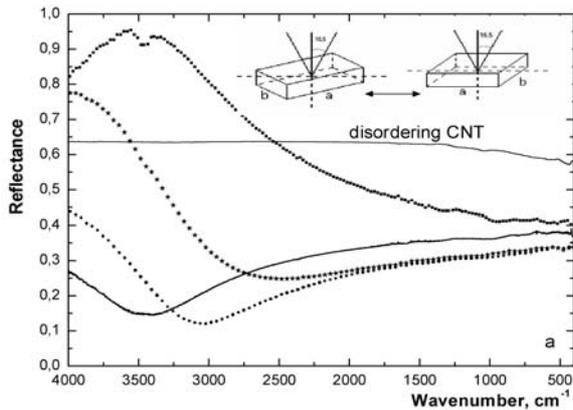

**Fig. 4.** Near-normal incidence IR reflectance spectra of the CNT carpet for different sample orientations: from the bottom layer of ordered carbon nanotubes (*a*), from side layers (*b*). Spectra of disordered powder of carbon nanotubes are presented for comparison.

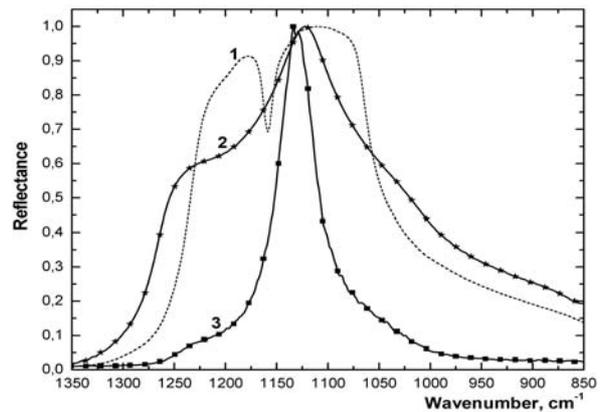

**Fig. 6.** Normalized IR reflectance spectra of single crystal of α-quartz (*1*), fused quartz (*2*) and opal crystal (*3*) in the region of stretching vibration (band of negative dielectric function).

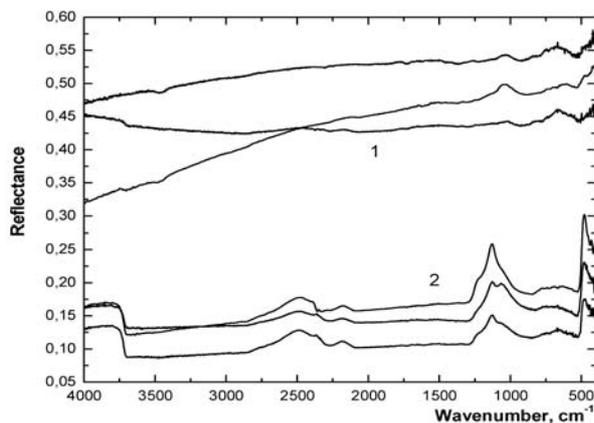

**Fig. 5.** Near-normal incidence IR reflectance spectra of the opal photonic crystal for different orientations of the sample: *1* – side layer (111), *2* – bottom layer.

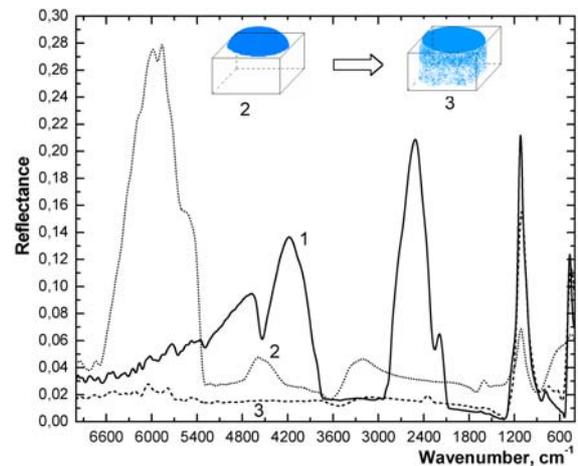

**Fig. 7.** IR reflectance spectra of initial opal after heating (*1*), opal with drop of the aqueous solution of poly-A just after drop deposition on the crystal surface (*2*) and dried film of poly-A on the opal surface (*3*).

It is known [10] that two types of bands in photonic crystals could exist: band gaps induced by Bragg scattering and other band gaps connected with negative permittivity. The angle dependence of Bragg scattering (shown in Fig. 3) is typically observed for opal photonic crystals in the visible range. However, some additional peaks appear in 6000-1000 cm$^{-1}$ region for opal, while they absent in solid silica (Fig. 5). We found a narrowing of the phonon band of opal in comparison with any solid crystal forms of silica. The halfwidth of the stretching vibration band (Fig. 6) for opal is 44 cm$^{-1}$ and less than those for fused quartz, namely, 88 cm$^{-1}$ and α-quartz, 102 cm$^{-1}$.

We wanted to use opal for enhanced IR absorption of biological molecules. So, we drop poly A water solution on opal surface. Saturation of opal pores by poly-A water solution causes to appearance of new bands in reflectance at 6000 and 4500 cm$^{-1}$ (Fig. 7, curve 3).

IR reflectance spectra of opal with drop of the aqueous solution of polyadenylic acid (poly-A) just after drop deposition on the opal surface (Fig. 7, curve 2) and dried film of poly-A on the opal surface (Fig. 7, curve 3) demonstrate only bands that can be assigned to the stretching vibration of opal (1300-900 cm$^{-1}$), deformation vibration (500-400 cm$^{-1}$). The bands of poly-A molecular vibrations are absent in this spectrum. That is why opal cannot be used for enhanced IR absorption of biological molecules. We suppose that elongated structures like cylinder, rod and strip create a forbidden band near 6000 cm$^{-1}$, like balls – in the 2000-5000 cm$^{-1}$. However, it needs a theoretical proof. Observed peculiarities in the IR spectra depend on a change of scale for the structure, very sensitive to disordering and do not depend on the molecular structure (phonon modes) of the material. Note that the Bragg gap is sensitive to disordering, whereas negative band induced by dielectric permittivity are insensitive to disordering.

In conclusion, a number of spectral peculiarities in the spectra of synthetic opals and ordered carbon nanotubes in the IR region were observed. These features indicate the narrowing of known phonon bands (negative ones) and appearance of new photon stop bands. A new theoretical approach should be developed for description of these processes. We suppose that the experimental data and ideas presented here could help to develop a theoretical model as well as open a new application of photonic crystals for optical filters, lenses, *etc*.

**Acknowledgements**

We thank Ukrainian Program – Nanostructured Systems, Nanomaterials, Nanotechnology, ("Bionanosystems" N0107U008449, 2007-2009) for financial support, Prof. Alberto Tagliaferro and his group from Torino Polithechnical Institute (Italy) for carpet nanotube production as well as for fruitful discussions and Dr.Sci. Viktor Stepkin from Institute of Physics, NAS of Ukraine (Kyiv) for SEM images of CNT carpet.